\documentclass[a4paper,amsmath,preprintnumbers,showpacs,twocolumn,prl,superscriptaddress, nofootinbib]{revtex4}
\usepackage{amssymb}
\usepackage{graphicx}
\usepackage{epstopdf}
\usepackage{bm}
\usepackage{epsfig}
\usepackage{graphics}
\usepackage{xspace}
\usepackage{amsfonts}
\usepackage{verbatim}
\usepackage{natbib}
\usepackage{color}

\graphicspath{{./eps/}}

\newcommand{\nn}{\nonumber}
\newcommand{\cut}{\textrm{cut}}

\begin{document}


\preprint{\font\fortssbx=cmssbx10 scaled \magstep2
\hbox to \hsize{
\hfill$\vcenter{
                \hbox{COEPP-MN-16-17}
                \hbox{MITP/16-054}
                }$}
}

\title{Fully Differential Higgs Pair Production in Association With a $W$ Boson at Next-to-Next-to-Leading Order in QCD}

\author{Hai Tao Li}\email{haitao.li@monash.edu}
\affiliation{ARC Centre of Excellence for Particle Physics at the Terascale, School of Physics and Astronomy, Monash University, Victoria, 3800 Australia}
\author{Jian Wang}\email{jian.wang@uni-mainz.de}
\affiliation{PRISMA Cluster of Excellence $\&$ Mainz Institute for Theoretical Physics,
Johannes Gutenberg University, D-55099 Mainz, Germany
\footnote{Present affiliation: Physik Department T31,  Technische Universit\"at M\"unchen, James-Franck-Stra\ss e~1,
D--85748 Garching, Germany. Email address: j.wang@tum.de}}




\begin{abstract}
To clarify the electroweak symmetry breaking mechanism, we need to probe the Higgs self-couplings,
which can be measured in Higgs pair productions. The associated production with a vector boson
is  special due to a clear tag in the final state.
We perform a fully differential next-to-next-to-leading-order calculation of the
Higgs pair production in association with a $W$ boson at hadron colliders,
and present numerical results at the 14 TeV LHC and a future 100 TeV hadron collider.

\end{abstract}


\maketitle


\emph{Introduction}:
It is of high importance to precisely measure the properties of the Higgs boson following its discovery in 2012 \cite{Aad:2012tfa,Chatrchyan:2012ufa}.
Present analyses  show that it is a spin-$0$ and CP-even particle with a mass of 125 GeV \cite{Khachatryan:2014kca}.
Its couplings with massive vector bosons have been measured to agree with the standard model (SM) expectations \cite{Aad:2013wqa,CMS-PAS-HIG-14-009}.
The couplings with heavy fermions, such as the top quark, the bottom quark and the $\tau$-lepton,
have also been determined in accordance with the SM \cite{Aad:2013wqa,CMS-PAS-HIG-14-009}.
The still unconfirmed properties are its self-couplings, which are crucial to clarify
the  electroweak symmetry breaking mechanism.
These couplings may be tested with the upcoming collision data at the LHC \cite{Baur:2003gp,Dolan:2012rv,Papaefstathiou:2012qe,Baglio:2012np,Barger:2013jfa,Barr:2013tda,Dolan:2013rja,Englert:2014uqa,
Liu:2014rva,ATL-PHYS-PUB-2014-019,CMS-PAS-FTR-15-002}
or a future 100 TeV hadron collider \cite{Yao:2013ika,Barr:2014sga,Li:2015yia,Azatov:2015oxa,Papaefstathiou:2015iba,Zhao:2016tai}.

\begin{figure}
  \includegraphics[width=0.9\linewidth]{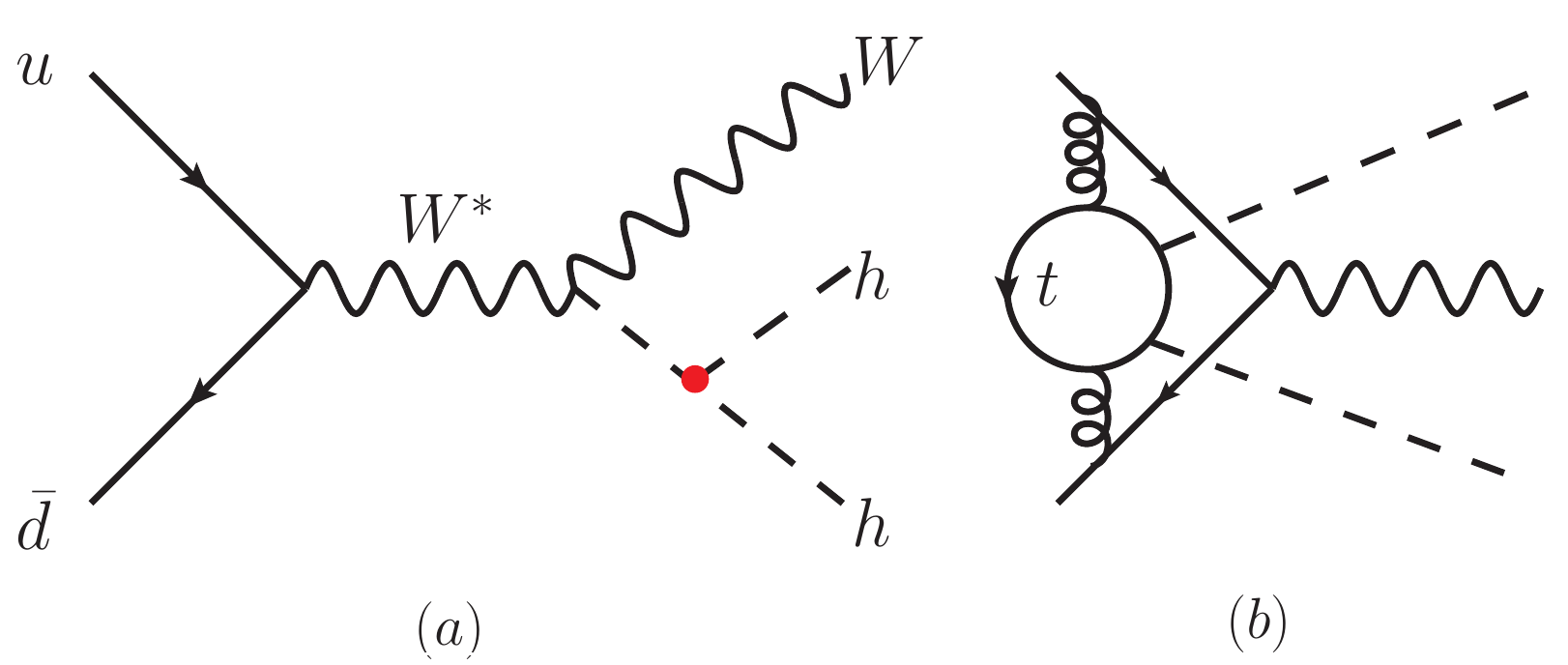}\\
  \caption{Selected Feynman diagrams for $Whh$ productions.
    $(a)$ is the LO diagram. $(b)$ is the NNLO diagram which we do not calculate in this work. }
  \label{fig:top_loop}
\end{figure}

Though it is possible to get some indications on the Higgs self-couplings from the virtual effects \cite{McCullough:2013rea},
the direct detection plays an indispensable role in probing these couplings.
The triple Higgs coupling can be measured by studying the Higgs pair productions at hadron colliders.
The dominant production channel is the gluon-gluon fusion which involves a top-quark loop.
The other channels, including the vector boson fusion, the vector boson associated production and the top quark pair associated production,
have relatively smaller cross sections.
One reason is that the phase space integration is smaller with more final-state particles.
However, the additional particles in the final state provide more handles on the signal so that
the backgrounds can be significantly suppressed.
Actually, the different channels have different characteristics, thus are complementary to each other and deserve discussion on the same footing.
In this work, we focus on the vector boson associated production channel, as shown in Fig.\ref{fig:top_loop}(a).
This channel is special for several reasons.
Since there is an associated vector boson that can serve as a characteristic tag,
one can select the events with the Higgs boson decaying to bottom quarks.
In this case, all the involved  Higgs couplings are not loop-induced,
avoiding the unknown  effects of virtual heavy particles.
Meanwhile, benefiting from the large branching fraction of the Higgs decay to bottom quarks,
the cross section of this channel is comparable to that of the gluon-gluon fusion production
and decay to $\gamma\gamma b\bar{b} $ \cite{Cao:2015oxx}.
Moreover, it depends on the value of the Higgs self-coupling in a different way from
the gluon-gluon fusion channel.
As a result, it is very sensitive to the Higgs self-coupling that is larger than the SM value \cite{Frederix:2014hta,Cao:2015oxx}.

The precise theoretical predictions are crucial for a proper interpretation of the experimental data.
The total cross section of the vector boson associated production  has been calculated up to next-to-next-to-leading order (NNLO)
in analogy to the Drell-Yan production  \cite{Baglio:2012np}.
However, in practice, experimental cuts are imposed on the final state.
It is not clear whether the NNLO corrections are the same over the full phase space.
Our aim in this work is to provide a fully differential NNLO calculation of the
Higgs pair production in association with a $W$ boson at hadron colliders.
This precise theoretical prediction can be used as a basic input
when analyzing the events in the future.

\emph{The method}:
As the QCD next-to-leading order (NLO) calculations of processes at hadron colliders become automatic,
the frontier has been upgraded to  fully differential calculations at NNLO accuracy.
In order to implement the kinematic cuts, the optical theorem, which has been used to calculate
the inclusive cross sections or decay rates beyond NLO, is not applicable.
One needs to deal with the virtual and real corrections separately,
which are individually divergent
\footnote{Here we talk about the infrared divergence.
The ultraviolet divergences in the virtual corrections are canceled by the standard renormalization procedure.}.
The divergences in virtual corrections can be obtained after integrating the loop momenta.
But it is harder to calculate the divergences in real corrections
because of the constraints (including on-shell condition and kinematic cuts) on the momenta of the final states.
Instead, it is essential to know the divergent behavior of the cross section near the infrared region,
including not only the coefficients of the singularities, like $1/\epsilon$ in $n=4-2\epsilon$ dimensional regularization scheme,
but also the finite contributions near the singularities.
The former,  arising when the energy of emitted gluons becomes vanishing
or the directions of some massless partons are collinear,
has been known up to 4-loop and 2-loop orders for the general massless \cite{Ahrens:2012qz}
and massive \cite{Becher:2009kw,Ferroglia:2009ep,Ferroglia:2009ii} parton scattering processes, respectively.
The latter depends on the way defined to approach the singularities in phase space.
At NLO, there is only one additional parton in real corrections with respect to the LO processes,
so it is clear to specify the phase space over which the integration of the scattering amplitude is divergent,
as proven in the FKS \cite{Frixione:1995ms} or the dipole subtraction methods \cite{Catani:1996vz}.
However, the task becomes very complicated at NNLO due to different combinations of the singular behavior of the
two additional partons; for example, see \cite{GehrmannDeRidder:2005cm}.

A way out of this complicity was invented by making use of the resummation method \cite{Catani:2007vq}.
The basic idea is that the cross section near the singular regions can be factorized
in terms of functions involving different energy scales individually.
Some of them describe the low-scale dynamics near the infrared divergent region,
which is independent of the high-scale dynamics in the hard collision and can be viewed as universal in this sense.
These functions include the structure
of the cross section near the singularities and can be calculated once for all.
Many of them have been obtained up to NNLO  so that a large number of processes
have been calculated at NNLO \cite{Catani:2007vq,Catani:2009sm,Catani:2011qz,Gao:2012ja,Cascioli:2014yka,Ferrera:2014lca,Gehrmann:2014fva,
Boughezal:2015dva,Boughezal:2015aha,Gaunt:2015pea,Boughezal:2015ded,Campbell:2016jau,Campbell:2016yrh,Grazzini:2016swo,Grazzini:2016ctr,
Boughezal:2016wmq,Berger:2016oht,deFlorian:2016uhr}.
In this work we extend these studies to the $2\to 3$ process of $pp\to Whh$,
which is important in probing the Higgs boson self-couplings at the LHC or future high-energy hadron colliders.

Since the final states are all color-singlets, additional partons,  at higher orders, can interact only with the initial states.
The system of the bosons $Whh$ has very small transverse momentum $q_T$ no matter whether the additional partons are soft or collinear
to the initial-state partons. This is  the region where the transverse momentum resummation is applied.
Using the well-known results in the resummation formalism,
one can predict the cross section with small transverse momentum, defined by $q_T<q_T^{\cut}$, up to NNLO in terms of universal functions.
Here $q_T^{\cut}$ is an intermediate cut-off parameter which will not appear in the final results.
Its value is chosen depending on the processes. See the following discussion for the process $pp\to Whh$.
The cross section with large transverse momentum  ($q_T>q_T^{\cut}$),  needed for NNLO corrections to $pp\to Whh$, is just the NLO corrections to $pp\to Whhj$,
which can be calculated using standard NLO subtraction method.

We first deal with the part with small $q_T$, and make use of the transverse momentum resummation
based on  soft-collinear effective theory (SCET)~\cite{Bauer:2000yr,Bauer:2001yt,Beneke:2002ph}.
Since the process of $pp\to Whh$ can be considered as a production of an off-shell $W$ boson and its decay to $Whh$,
the cross section for $Whh$ production in the small $q_T$ region, analogy to Drell-Yan production,
can be written as~\cite{Becher:2011xn}
\begin{multline}
  \frac{d\sigma}{dq_T^2 dydM^2} =\frac{1}{2 sM^2} \sum_{i,j=q\bar{q}g}\int_{\zeta_1}^1 \frac{dz_1}{z_1}
      \int_{\zeta_2}^1 \frac{dz_2}{z_2}\int d\Phi_3
     \\
      \times
     f_{i/N_1}(\zeta_1/z_1,\mu)f_{j/N_2}(\zeta_2/z_2,\mu) H_{q\bar{q}}(M,\mu)
     \\
     \times  C_{q\bar{q}\leftarrow i j}(z_1,z_2,q_T, M,\mu)
+\mathcal{O}\left(\frac{q_T^2}{M^2}\right)~
\label{eq:dsigma}
\end{multline}
where $q_T$, $M$ and $y$ are transverse momentum, invariant mass and rapidity of $Whh$ system.
And $ d\Phi_3$ represents three-body phase space at Born level.
$ f_{i/N}(x,\mu)$ denotes the parton distribution function (PDF).
The hard function $H_{q\bar{q}}(M,\mu) $ contains the contribution from high-scale interactions, independent of $q_T$, and
is extracted by matching the (axial) vector current in full QCD onto an effective current built out of operators in SCET,
whose two-loop results can be obtained from hard functions of the Drell-Yan process~\cite{Becher:2006mr}.
Starting from two-loop, there is an additional contribution from diagrams with the Higgs boson(s) emitted from a virtual top-quark loop inside a gluon propagator, as shown in Fig.\ref{fig:top_loop}(b).
This kind of contribution also occurs in $pp\to Wh$ production, and it is found to be less than $2.1\%$ of the LO cross section, as shown in Fig.6(c) of Ref.\cite{Brein:2011vx}.
If we assume a similar correction in $pp\to Whh$ production, it is much less than the NNLO corrections
we have considered in this work, which is about $25\%$; see Fig.\ref{fig:sqrts} below.
The exact evaluation requires the calculation of the multi-scale five-point non-planar two-loop diagrams,
which is left to future work.
Therefore, we do not include this kind of contribution in the present work.
All the $q_T$ dependent terms are contained in the collinear kernel
\begin{align} \label{eq:c_scet}
    & C_{q\bar{q}\leftarrow i j}(z_1,z_2,q_T, M,\mu)
     = \frac{1}{4\pi}\int d^2x_\perp e^{- i x_\perp \cdot q_\perp} \nn \\
    &  \left(\frac{x_T^2 M^2}{b_0^2}\right)^{F_{q\bar{q}}(x_T^2, \mu)}
     I_{q\leftarrow i}(z_1,L_\perp, \alpha_s) I_{\bar{q}\leftarrow j}(z_2,L_\perp, \alpha_s)
\end{align}
with $  L_\perp = \ln\frac{x_T^2\mu^2}{b_0^2}$ and $  b_0 = 2 e^{-\gamma_E}$.
The  function $F_{q\bar{q}}(x_T^2, \mu)$ arises from the effect of collinear anomaly and
plays a special role in relating the traditional transverse-momentum resummation formalism and that in SCET~\citep{Becher:2011xn}.
The kernel $I_{q\leftarrow i}$ describes the evolution of a parton $i$ to $q$.
Their two-loop results  have been obtained recently in Refs.~\cite{Gehrmann:2012ze,Gehrmann:2014yya}~.
With all the NNLO ingredients available  it is straight forward to perform the integration of $q_T$ from $0$ to $q_T^{\cut}$ in Eq.(\ref{eq:dsigma}).

Next we move to the cross section with large $q_T$.
As mentioned before, this part amounts to the NLO corrections to $Whhj$ production that
can be tackled with standard NLO techniques.
The only different point is that we do not apply any jet algorithm on the final state.
It is only in this case that the combination of phase spaces of $pp \to Whhj$ at NLO with large $q_T$ and
$pp \to Whh$ at NNLO with small $q_T$ can recover the whole phase space of $pp\to Whh$ at NNLO.
One does not need to worry about the problem of infrared divergences due to the lack of a jet algorithm.
The reason is that the basic constraint $q_T> q_T^{\cut}$ prevents the momentum of the jet in $pp \to Whhj$
to be arbitrarily soft or collinear to the initial-state partons.
In practice, the only problem is that the numerical result may converge slowly if $q_T^{\cut}$
is chosen to be too small.
In this work, we use MadGraph5\_aMC@NLO \cite{Alwall:2014hca} to calculate the NLO corrections automatically.
Actually, this is one of the advantages when using $q_T$ subtraction, i.e.,
the present tools and programs of NLO calculations can be utilized without any substantial change.

Combining the two parts together, we obtain the NNLO differential cross section of the process $pp\to Whh$
\begin{multline}\label{eq:main}
   \frac{d\sigma_{Whh}}{d\Phi_3dy}\Big\vert_{\textrm{NNLO}} =
   \\
   \underbrace{\int_0^{q_T^{\cut}} dq_T \frac{d\sigma_{Whh}}{d\Phi_3 dy dq_T} }_\text{SCET}
   + \underbrace{\int^{q_T^{\max}}_{q_T^{\cut}} dq_T \frac{d\sigma_{Whhj}}{d\Phi_3dydq_T}}_\text{MadGraph5\_aMC@NLO}
\end{multline}
where $q_T^{\max}$ is set by the partonic center-of-mass energy and the invariant mass of $Whh$.

\begin{figure}
  \includegraphics[width=0.9\linewidth]{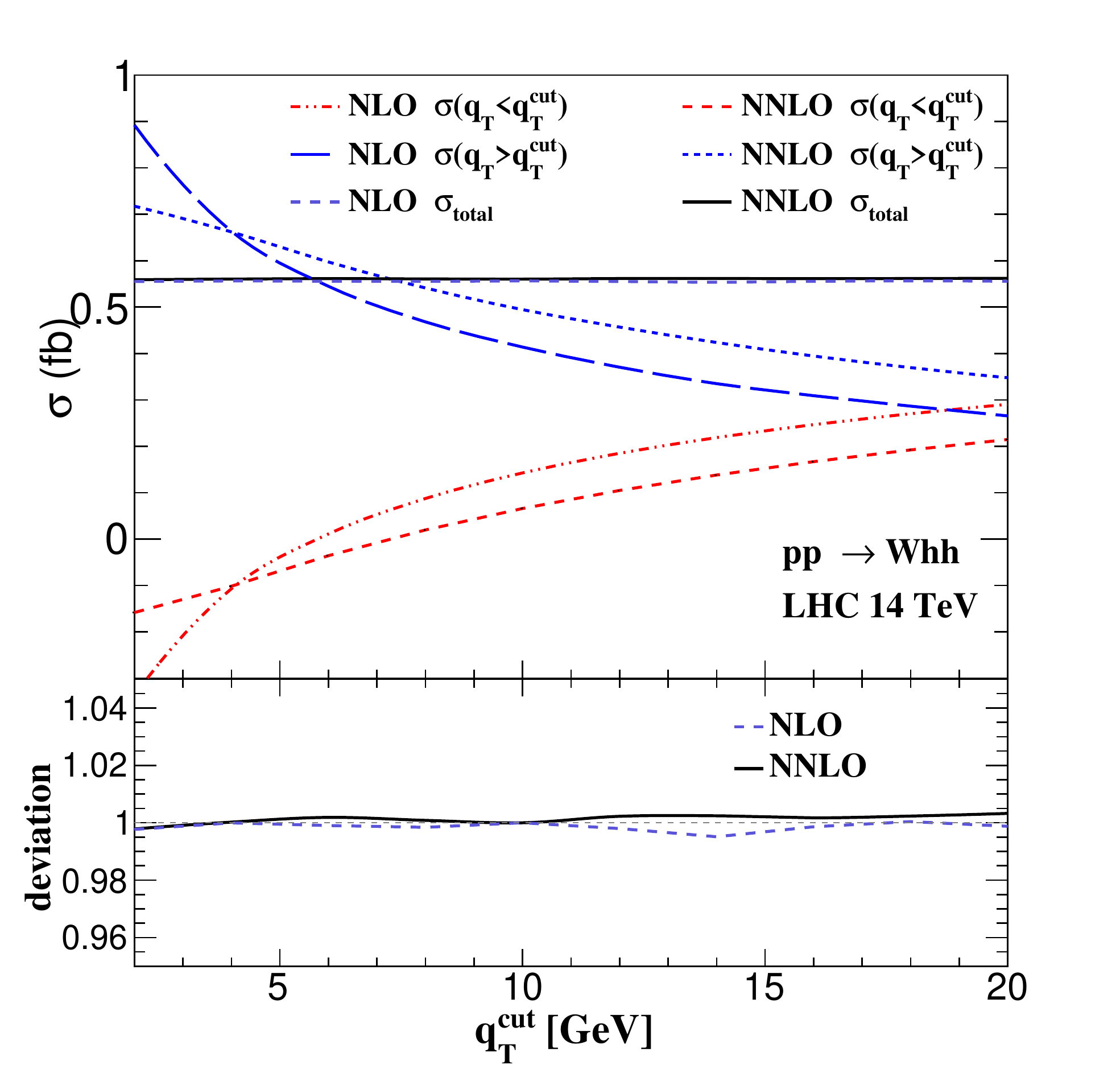}\\
  \caption{The total cross sections of $pp\to Whh$ production at NLO and NNLO in QCD.
    In the bottom plot, the deviation is defined as $\sigma(q_T^{\rm cut})/\sigma(q_T^{\rm cut}=10~{\rm GeV})$~. }
  \label{fig:qTcut}
\end{figure}

\emph{Numerical results}:
We now present the numerical results for $Whh$ (including $W^+hh$ and $W^-hh$) production at the 14 TeV LHC and a future 100 TeV hadron collider.
We use CT14 PDF set  and associated strong coupling evaluated at each corresponding order throughout our calculation \cite{Dulat:2015mca}.
The relevant non-vanishing CKM matrix elements are
$V_{ud}=0.97425,~ V_{us}=0.2253,~ V_{ub}=4.13\times 10^{-3},~
V_{cd}=0.225,~ V_{cs}=0.986,~ V_{cb}=4.11\times 10^{-2}$ \cite{Agashe:2014kda}.
The other input parameters are chosen as:
\begin{align}
& M_W = 80.419 ~\textrm{GeV},
\quad m_h = 125 ~\textrm{GeV},
\quad \sin^2 \theta_W = 0.222 \nn \\
& \hspace{2cm} \alpha = \frac{1}{132.507},
\quad \lambda_{hhh}^{\rm SM} = \frac{m_h^2}{2v}.
\end{align}
The default factorization scale $\mu_F$ and renormalization scale $\mu_R$ are set equal to $M$
in order to avoid possible large logarithms.
As shown in Eq.(\ref{eq:main}), the two contributions on the right-hand side
depend on the cut-off parameter $q_T^{\cut}$ individually, though their sum on the left-hand side
is independent of it.
Therefore, it is crucial to first check this feature of the method numerically.
In Fig.{\ref{fig:qTcut}}, we show the total cross sections of  $pp\to Whh$ production at NLO and NNLO in QCD
as a function of $q_T^{\cut}$.
One can see that the total cross sections are almost unchanged as $q_T^{\cut}$ varies from 2 GeV to 20 GeV,
though the individual parts $\sigma(q_T<q_T^{\rm cut})$ and $\sigma(q_T>q_T^{\rm cut})$ depend on the cutoff strongly.
Notice that the typical scale of this process is about $M\sim 500$~GeV.
Therefore, the power corrections in this method are about $(q_T^{\rm cut}/M)^2\sim 0.04\%$ for the choice of $q_T^{\rm cut}=10$~GeV,
which can be safely neglected.
In the following discussion  we choose $q_T^{\rm cut}$ at 10 GeV.
As a cross check, we have compared our NLO total and differential cross section obtained by Eq.(\ref{eq:main})
with that by the standard NLO program MadGraph5\_aMC@NLO \cite{Alwall:2014hca} and found good agreement.

\begin{figure}
    \includegraphics[width=0.9\linewidth]{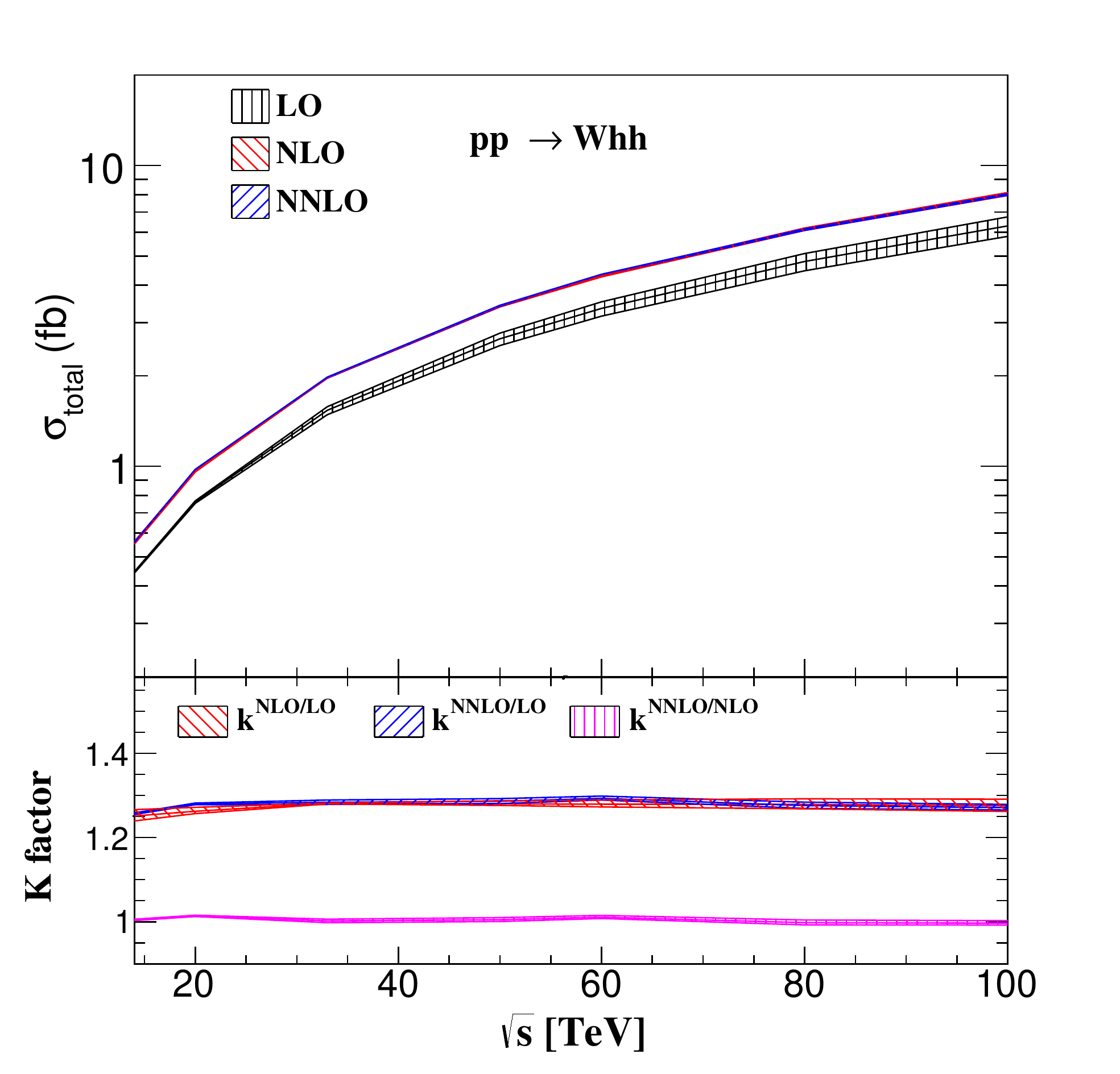}\\
  \caption{The total cross sections of $pp\to Whh$ production as a function of the collision energy.
The bands denote the scale uncertainties when varying $\mu=\mu_F=\mu_R$ by a factor of two. }
  \label{fig:sqrts}
\end{figure}

Then we report the total cross sections at different collision energies in Fig.{\ref{fig:sqrts}}.
One can see that the cross sections increase quickly with the increasing of collision energy.
The LO results suffer from large scale uncertainties when the collision energy is large.
In contrast, the NLO and NNLO results have very small scale uncertainties, and thus provide more precise predictions.
The $K$-factors, defined as the ratio of higher-order results over the lower-order ones,
indicate the effects of higher-order corrections.
The NLO and NNLO $K$-factors are nearly the same, both around $1.25\sim 1.3$ when the collision energy changes from 14 TeV to 100 TeV.
By adopting the same PDF sets, we also reproduce the total cross sections given in the literature \cite{Baglio:2012np},
which can be considered as a strong check of our calculation.

\begin{figure}
  \includegraphics[width=0.45\linewidth]{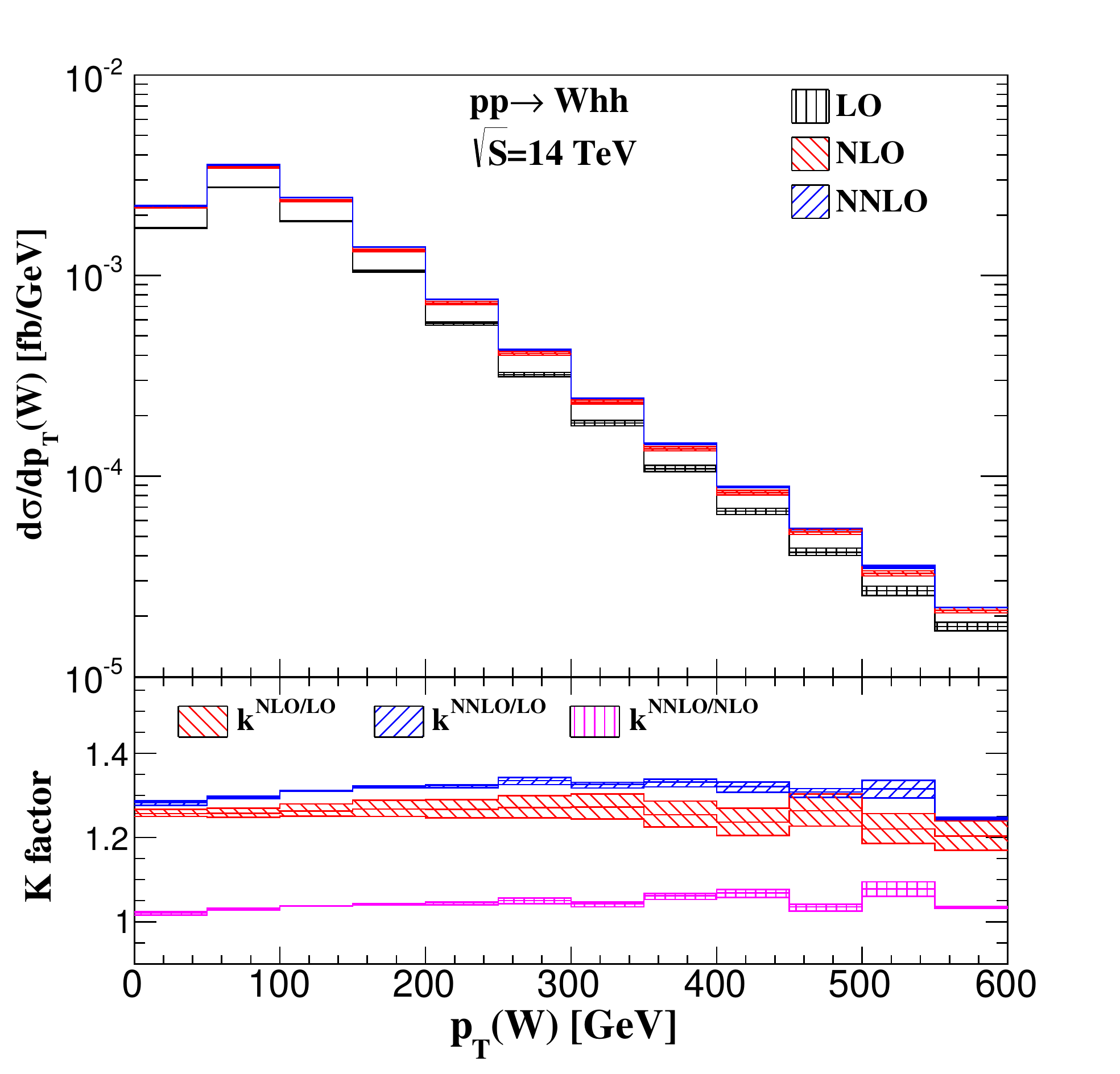}
    \includegraphics[width=0.45\linewidth]{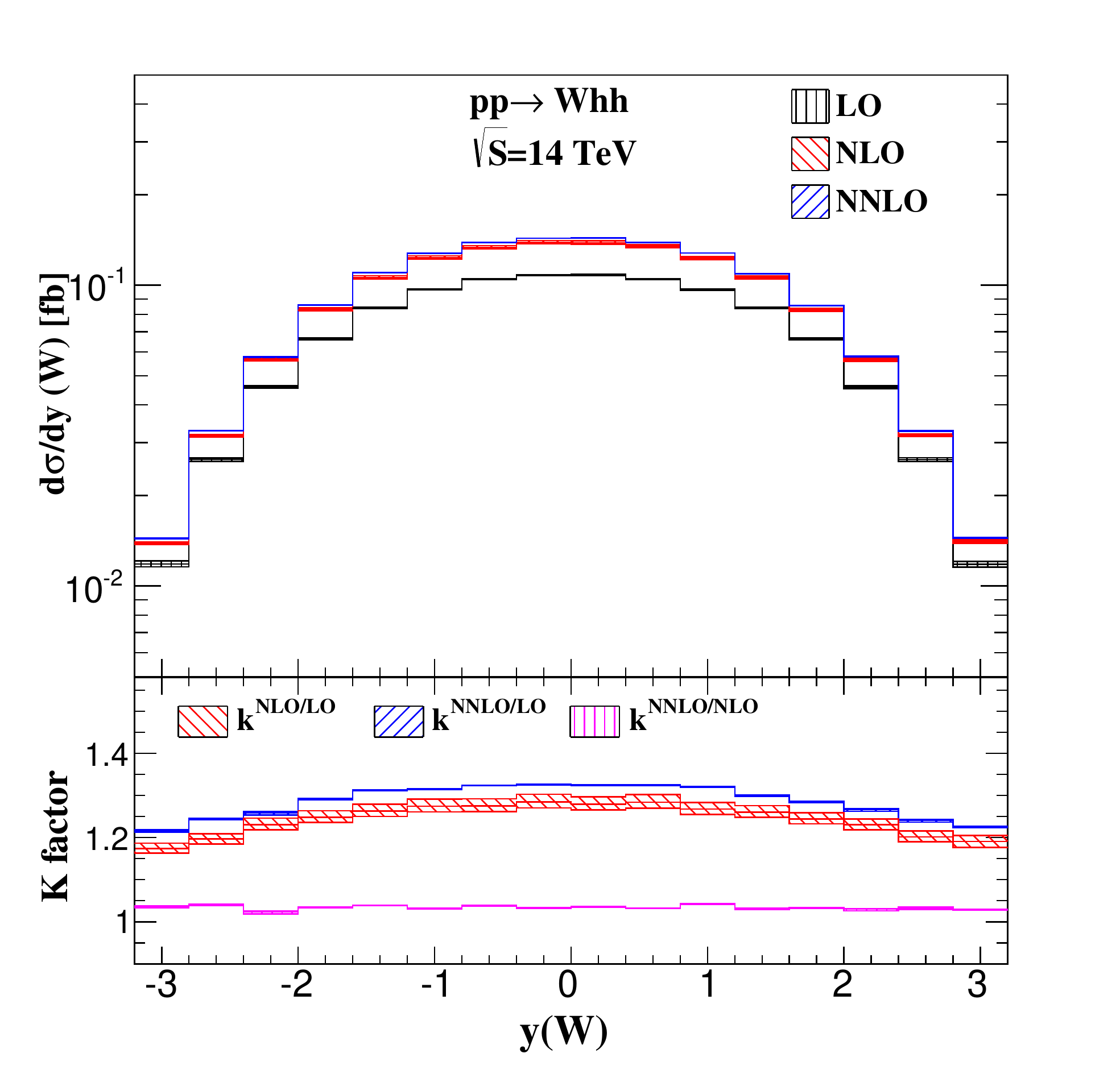}
    \includegraphics[width=0.45\linewidth]{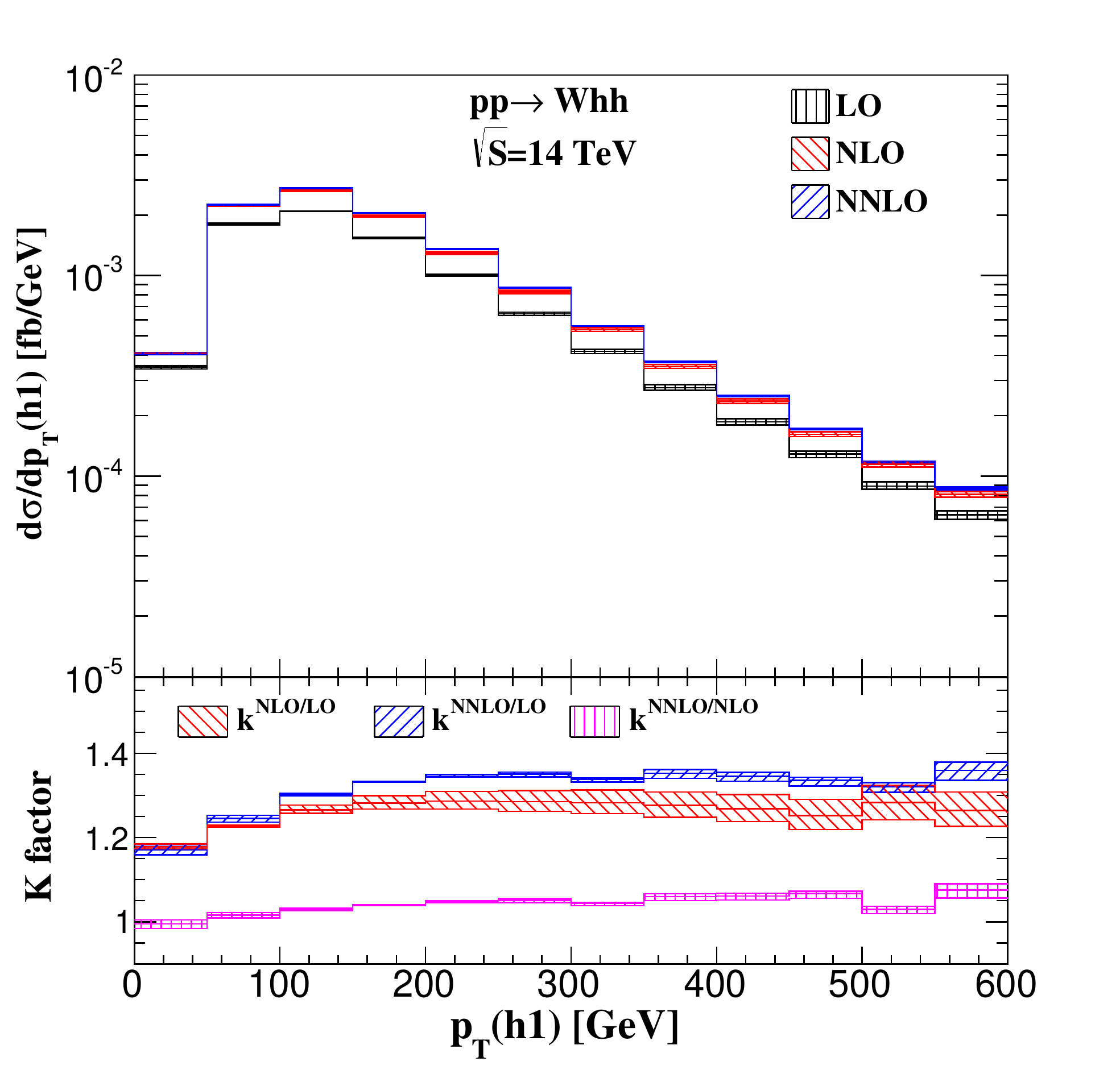}
    \includegraphics[width=0.45\linewidth]{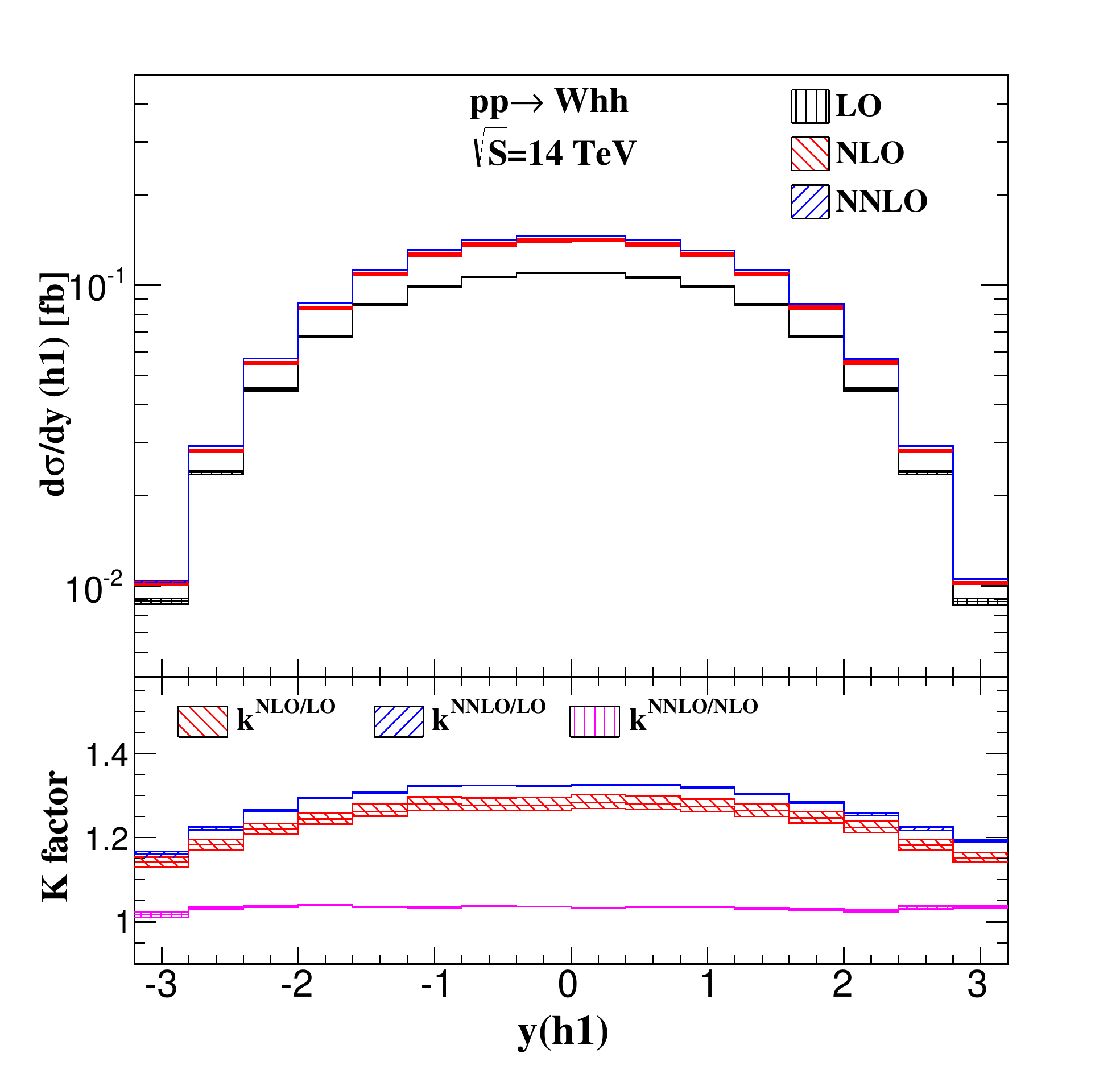}
    \includegraphics[width=0.45\linewidth]{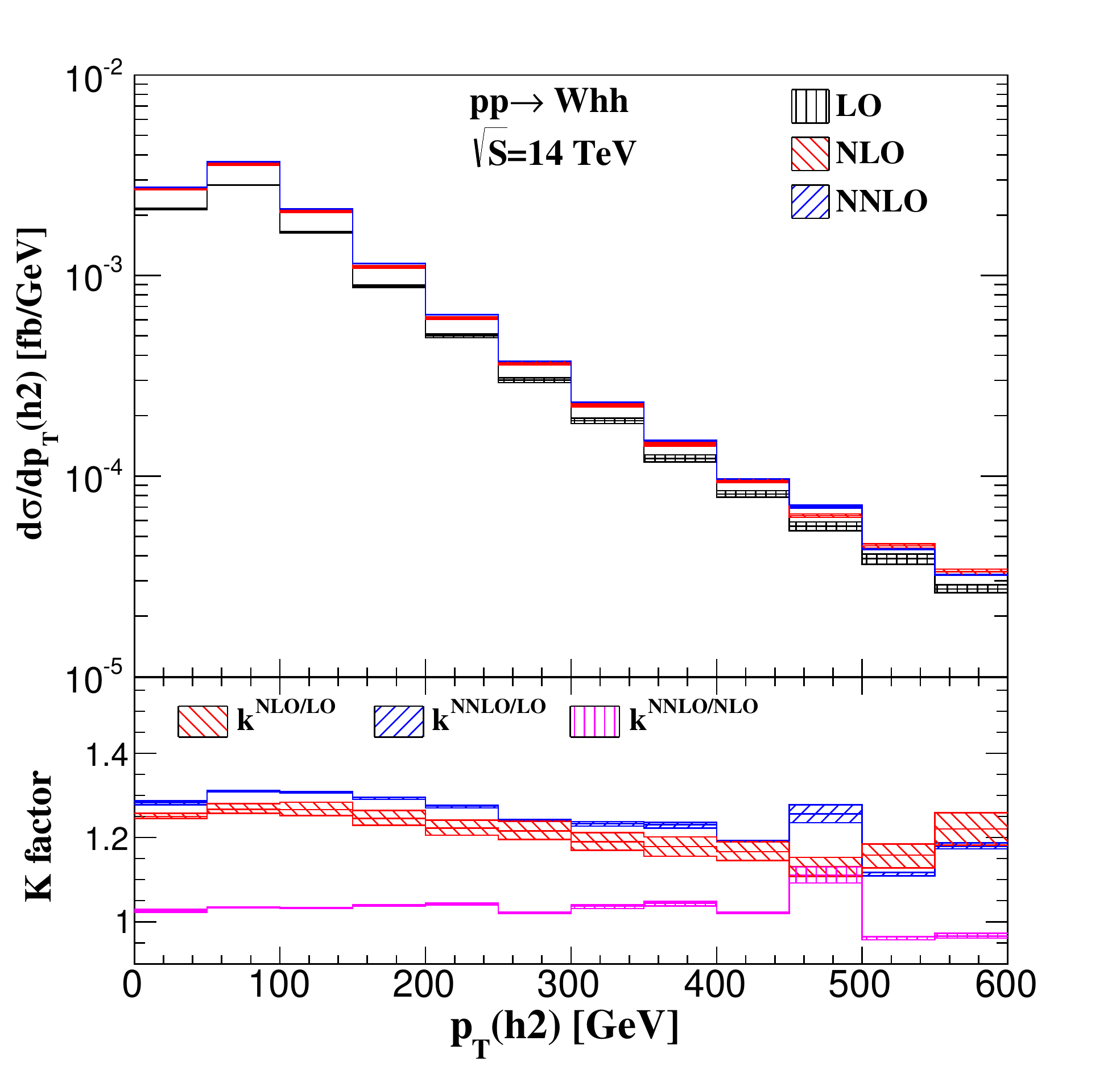}
    \includegraphics[width=0.45\linewidth]{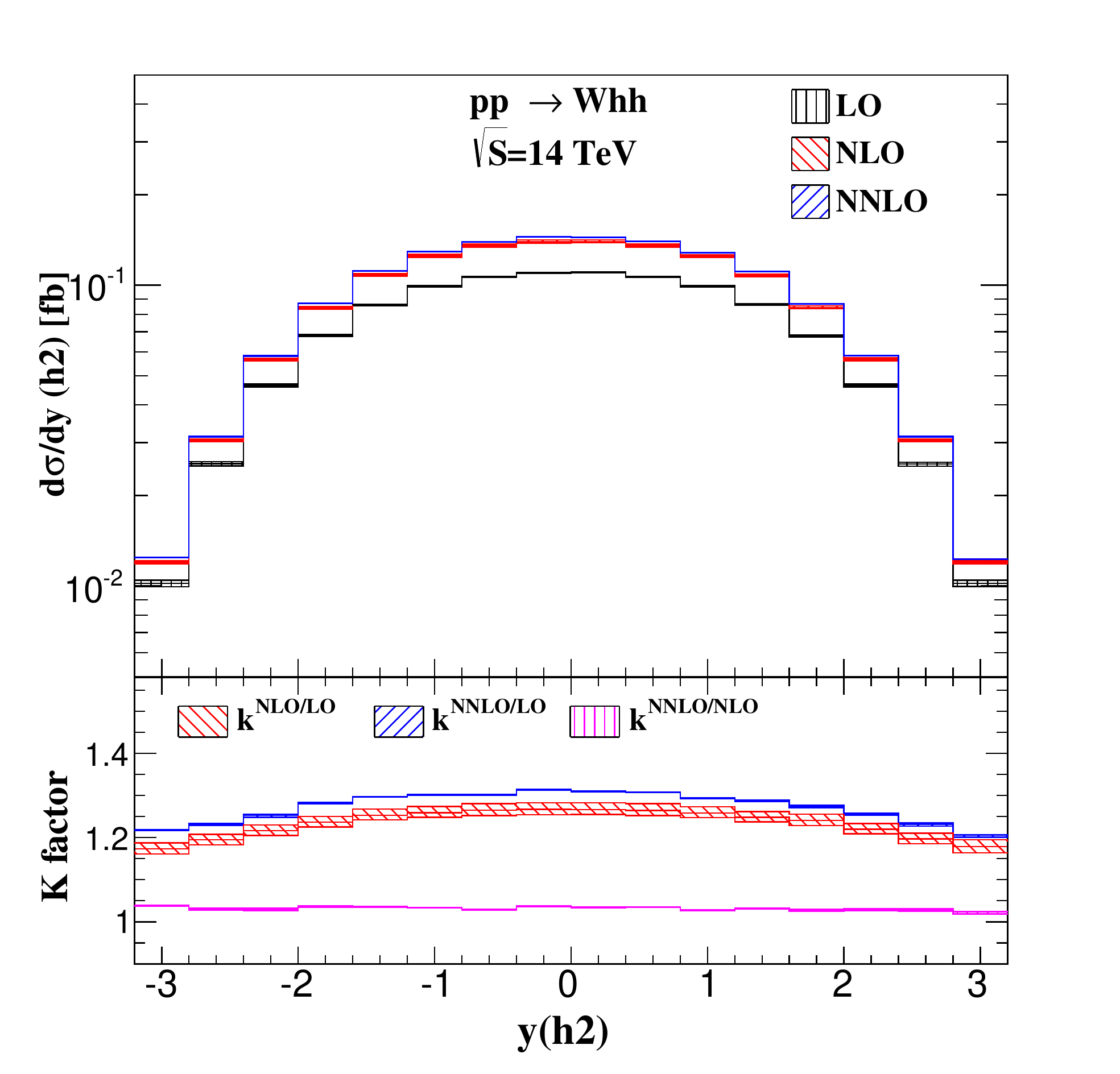}\\
  \caption{The kinematic distributions of $pp\to Whh$ production at the 14 TeV LHC.
  $h1(h2)$ denotes the Higgs boson with larger (smaller) transverse momentum. }
  \label{fig:kin}
\end{figure}

\begin{figure}
  \includegraphics[width=0.45\linewidth]{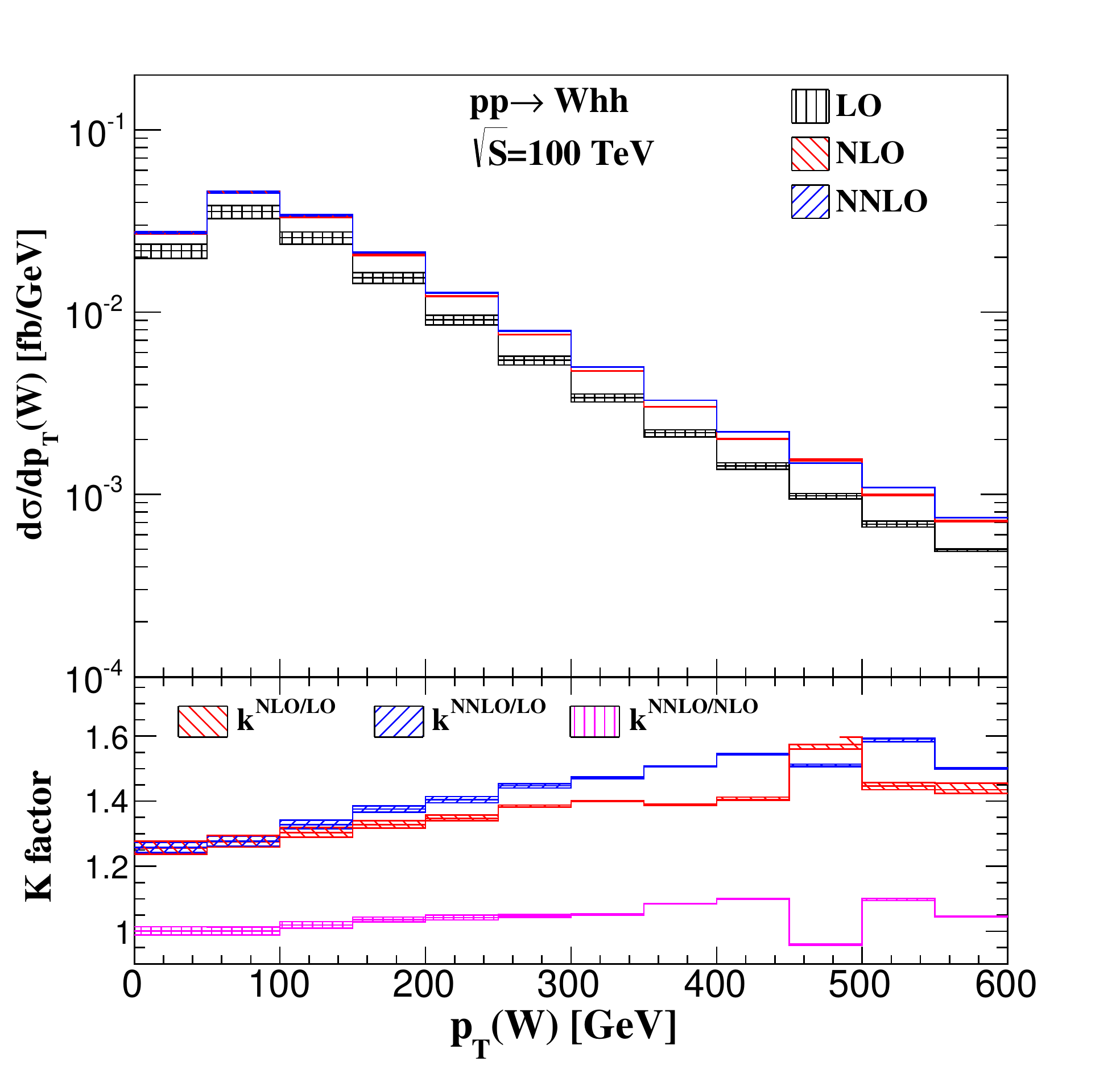}
    \includegraphics[width=0.45\linewidth]{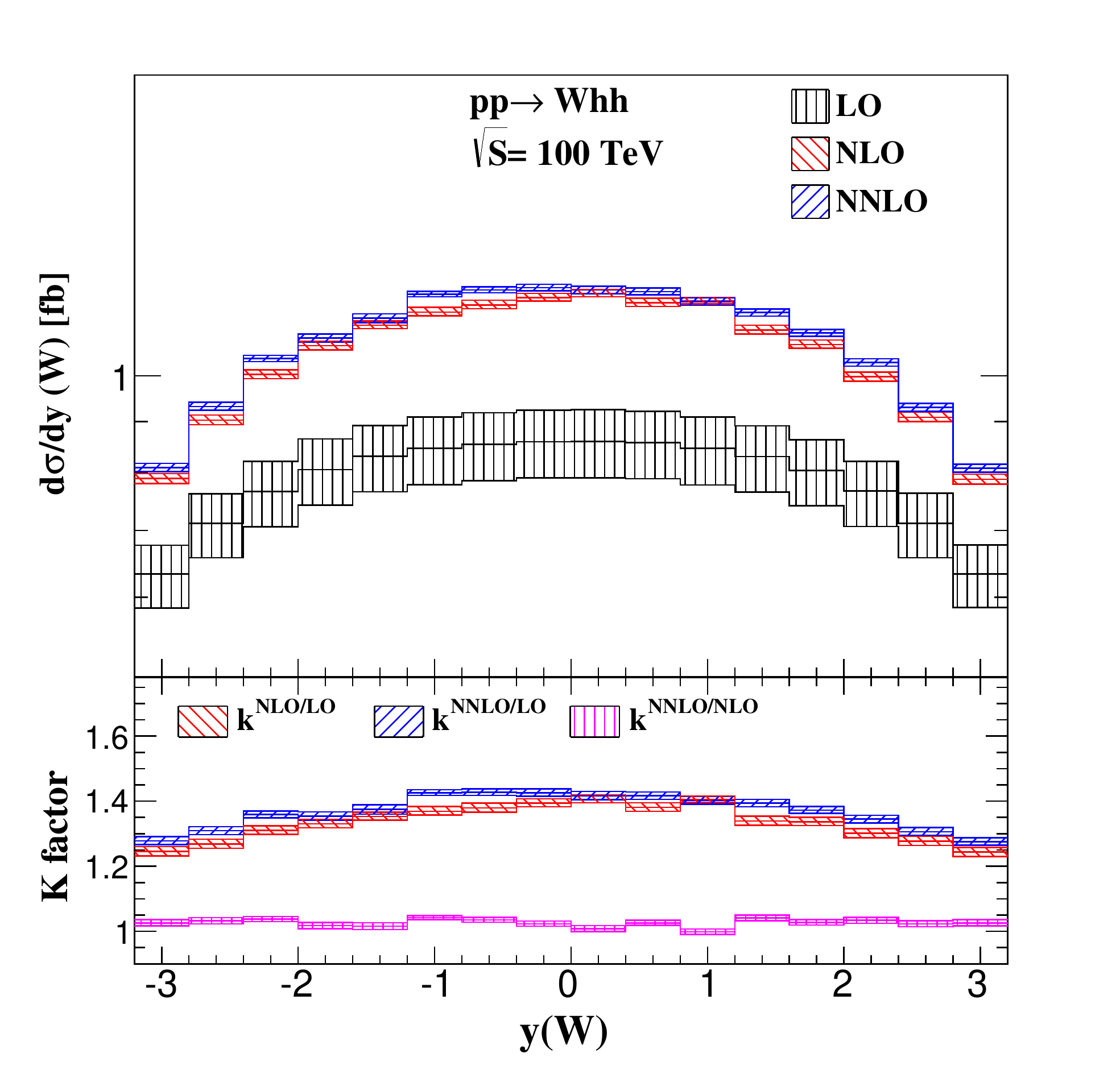}
    \includegraphics[width=0.45\linewidth]{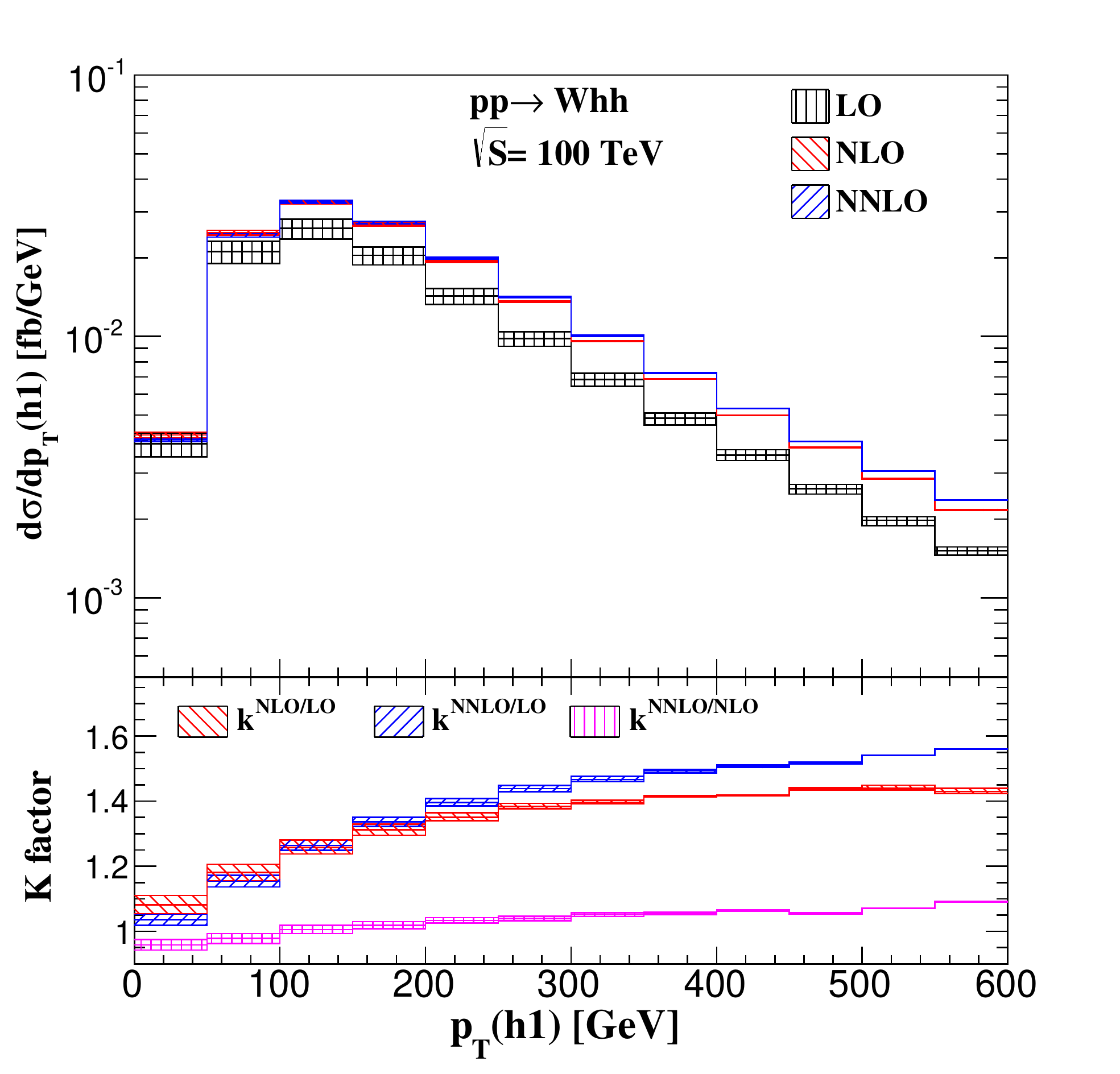}
    \includegraphics[width=0.45\linewidth]{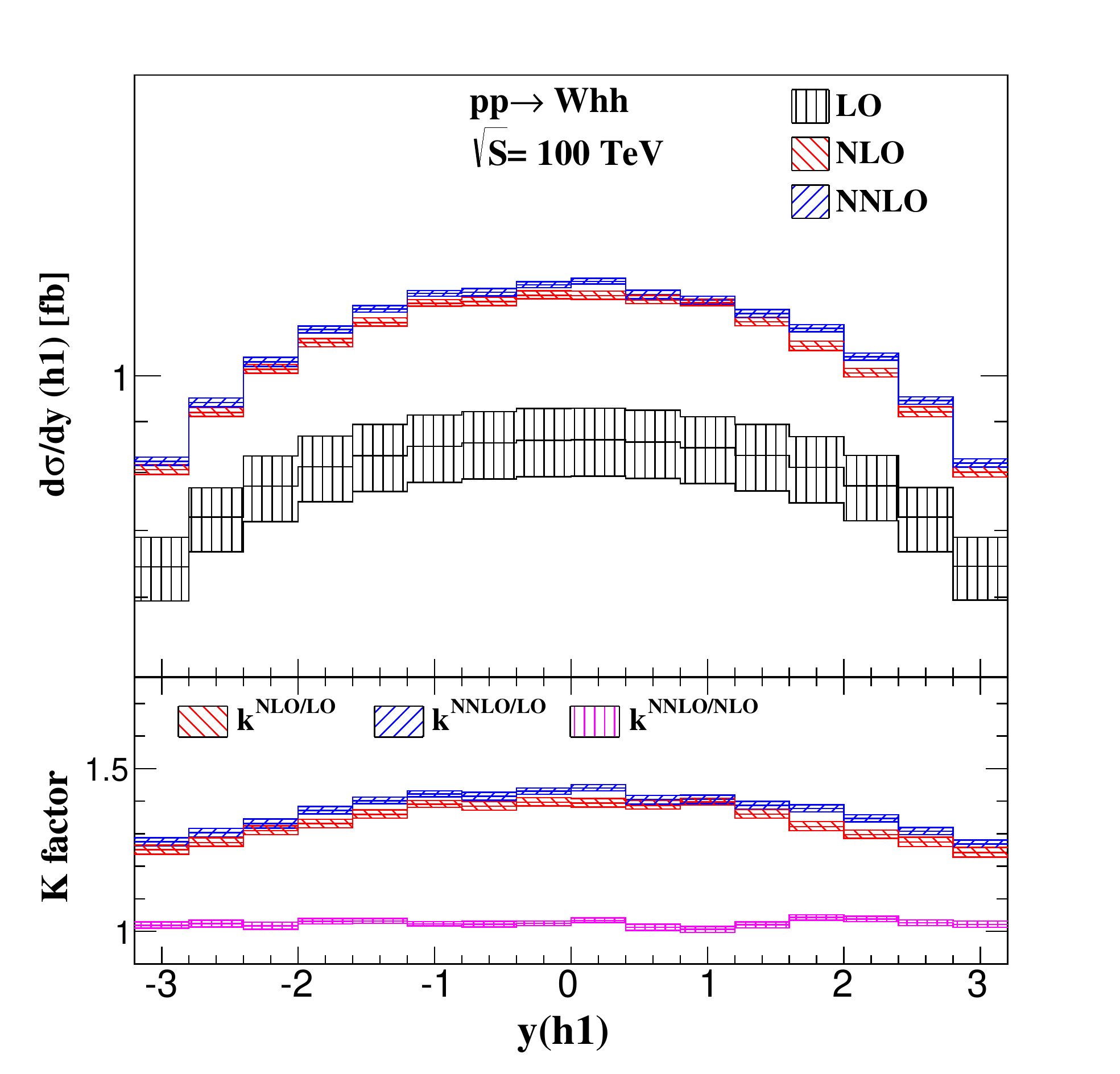}
    \includegraphics[width=0.45\linewidth]{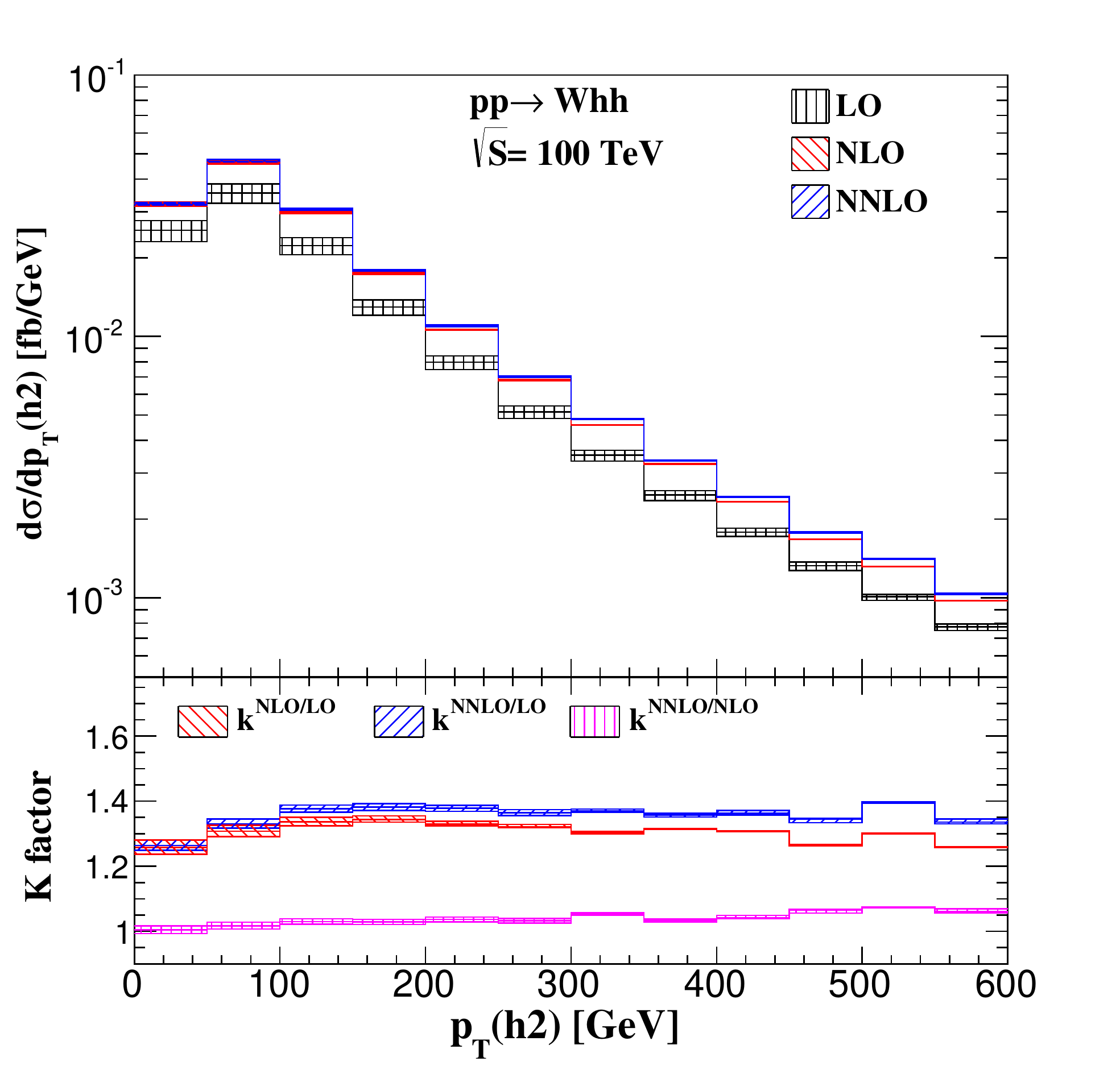}
    \includegraphics[width=0.45\linewidth]{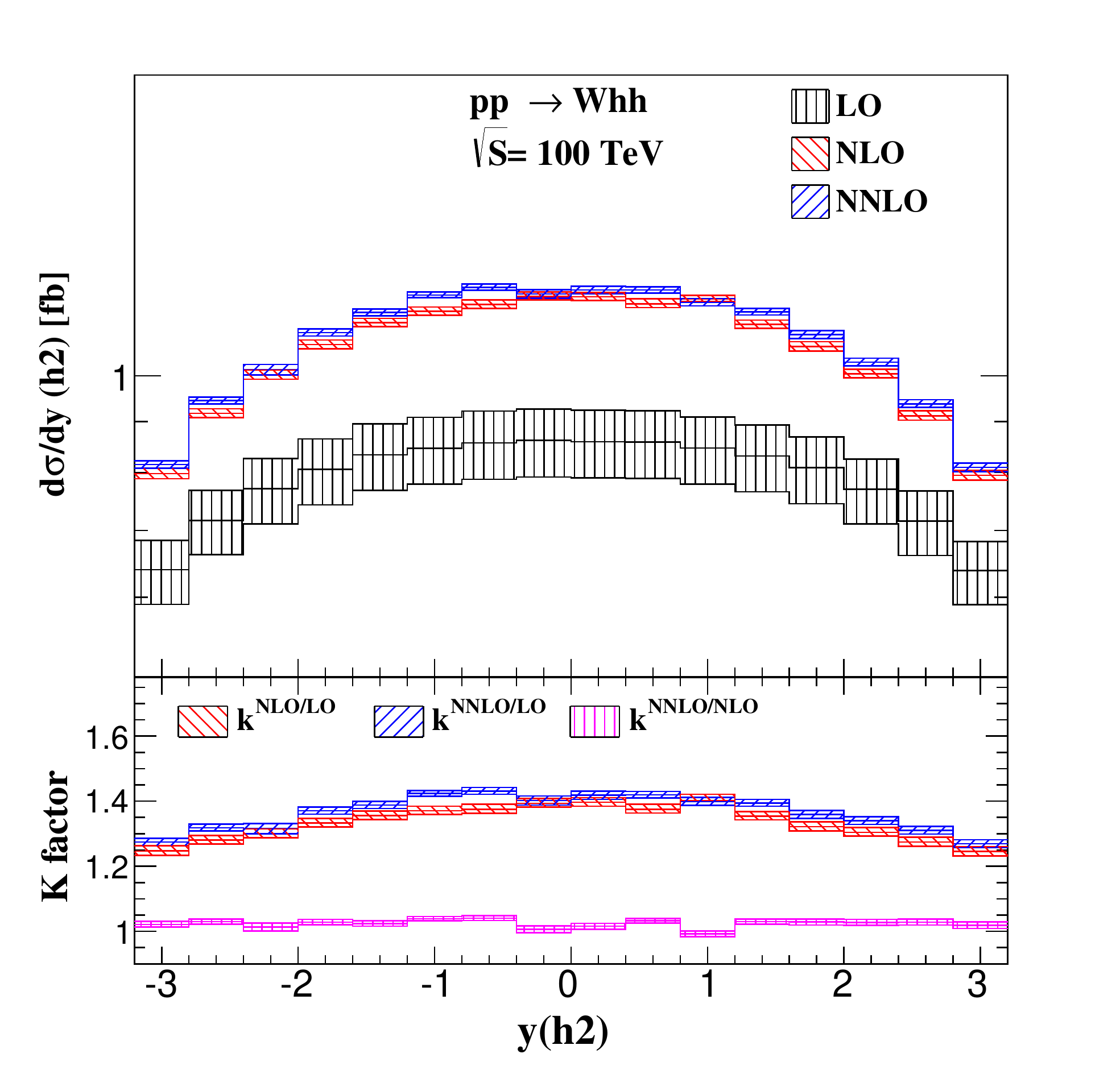}\\
  \caption{The kinematic distributions of $pp\to Whh$ production at a future 100 TeV hadron collider.
   $h1(h2)$ denotes the Higgs boson with larger (smaller) transverse momentum.  }
  \label{fig:kin100}
\end{figure}

Now we present the kinematic distributions  of $pp\to Whh$ production at the LHC in Fig.{\ref{fig:kin}}.
It can be seen that the NLO and NNLO $K$-factors are very similar in various distributions.
They change slightly for the $W$ boson transverse momentum but  grow with the increasing of the leading Higgs boson transverse momentum.
They have relatively larger values in the central rapidity region of both the $W$ and Higgs bosons
with respect to the forward and backward region.
In all distributions, the scale uncertainties have been reduced at NNLO,
which makes the prediction more reliable.

The results at a 100 TeV hadron collider are shown in  Fig.{\ref{fig:kin100}}.
One can see that the NNLO corrections enhance the cross section significantly
in the large transverse momentum region.
To make it more clear, we apply the following kinematic cuts
\begin{align}
& p_T(W)>200~\textrm{GeV},\quad |y(W)|<2.4, \nn \\
& p_T(h)>200~\textrm{GeV},\quad |y(h)|<2.4
\end{align}
to select the highly boosted events.
The cross sections under these cuts are shown in Table~\ref{tab:sigma}.

\begin{center}
\begin{table}
\centering
\begin{tabular}{c| c| c}
\hline\hline
$\sigma ~[fb]$ & boosted region & jet veto
\\
\hline
LO & 0.271$^{+3.0\%}_{-3.5\%}$  & 6.30$^{+7.1\%}_{-7.7\%}$
\\
NLO &  0.360$^{+0.5\%}_{-0.1\%}$  & 3.75$^{+6.4\%}_{-5.8\%}$
\\
NNLO & 0.382$^{+0.7\%}_{-0.5\%}$  & 3.01$^{+2.7\%}_{-2.4\%}$
\\
\hline
$K^{\rm NLO/LO}$ & 1.33  & 0.60
\\
$K^{\rm NNLO/LO}$ & 1.41  & 0.48
\\
$K^{\rm NNLO/NLO}$ & 1.06  & 0.81
\\
\hline\hline
\end{tabular}
\caption{\label{tab:sigma} The cross sections at a 100 TeV hadron collider after kinematic cuts. The applied cuts are
described in the text.}
\end{table}
\end{center}

In practice, for a process with a $W$ boson in the final state, applying a jet veto can suppress the background from top quarks substantially.
To investigate this kind of effect, we calculate the vetoed cross section
by discarding the events containing any jet with $p_T({\rm jet})>30$~GeV and $|\eta(\rm jet)|<3.5$.
Here jets are constructed by the anti-$k_t$ algorithm \cite{Cacciari:2008gp} with a radius of $R=0.7$.
One should apply the jet algorithm in such a way that the dependence of the cross section of $pp\to Whhj$
on $q_T^{\cut}$ is not affected.
In the part with $q_T>q_T^{\cut}$, the momenta of final-state particles are generated when performing phase space integration,
and thus there is no ambiguity to apply the jet algorithm.
But in the part with $q_T<q_T^{\cut}$, the momenta of emitted partons have been already integrated in the beam function,
in which only the total transverse momentum of the emitted partons is constrained, ignoring any effect from the jet algorithm.
In principle, one would need another kind of beam function in order to predict the jet-vetoed cross sections.
In practice, if the jet veto is much larger than $q_T^{\cut}$, the difference between the two kinds of beam functions at NNLO
are those emissions in which both of the two emitted partons must have transverse momenta around or larger than the jet veto and
move in nearly the back-to-back directions in the transverse plane with their sum less than  $q_T^{\cut}$.
The contributions of these emissions to the beam function are strongly suppressed. In our calculation of jet-vetoed cross sections,
we take $q_T^{\cut}=6$ GeV, much less than the jet veto.
We have also tried even smaller value of $q_T^{\cut}=2$ GeV,
finding that the numerical results nearly unchanged.

In Table~\ref{tab:sigma} we report the corresponding cross sections  at a 100 TeV hadron collider after a jet veto.
We first notice that the effect of jet veto starts from NLO and that the higher-order corrections are very sizable.
The NLO cross sections are only $60\%$ of the LO ones.
And the NNLO corrections decrease the NLO cross sections further by $19\%$.
Then we observe that the scale uncertainties are still large at NLO.
However, after including NNLO corrections, the scale uncertainties become significantly reduced.

\begin{figure}
  \includegraphics[width=0.6\linewidth]{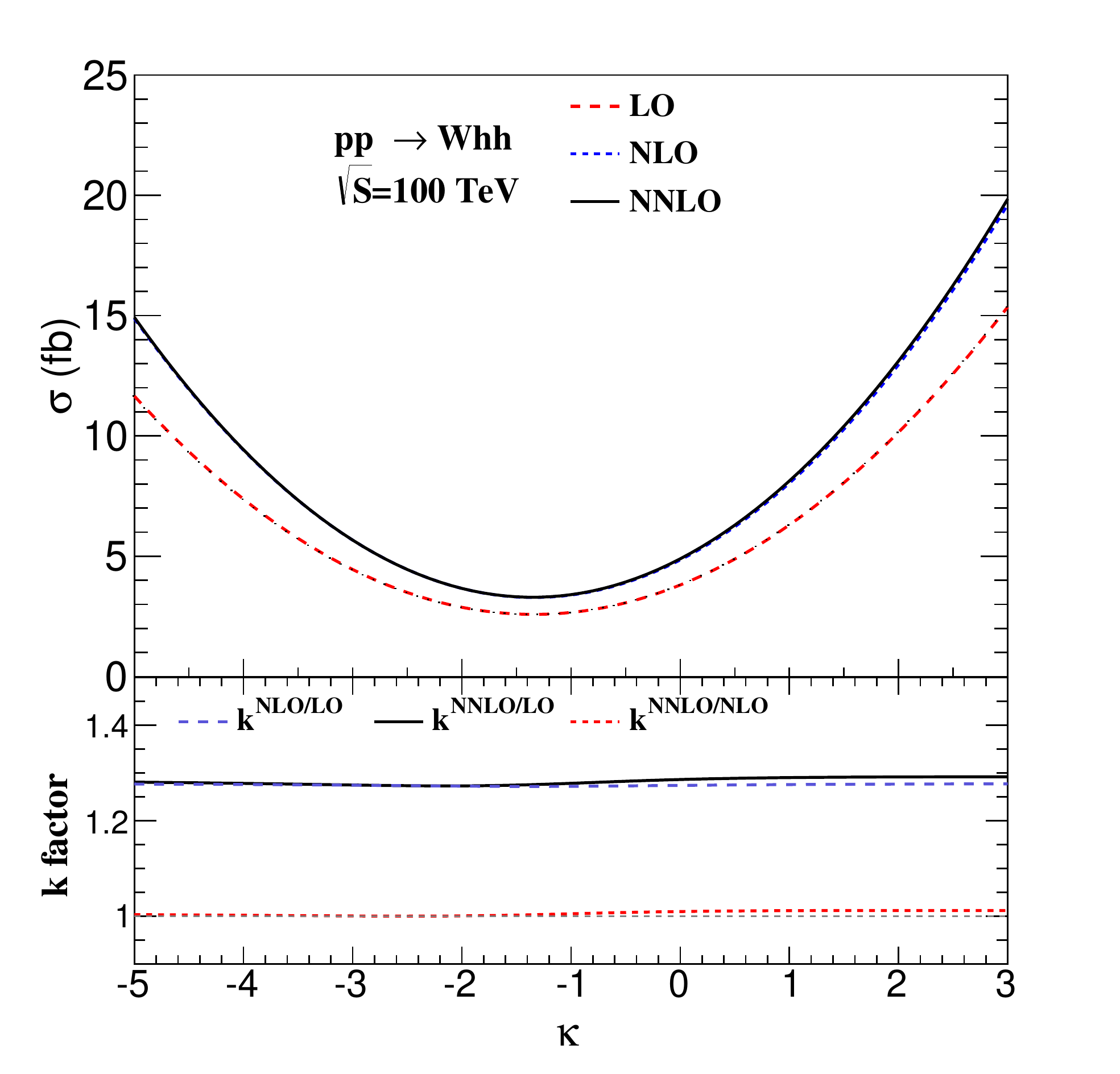}
  \caption{ The production cross sections as a function of $\kappa$ with $\sqrt{S}=100$ TeV. }
  \label{fig:kappa_dep}
\end{figure}

\begin{figure}
  \includegraphics[width=0.45\linewidth]{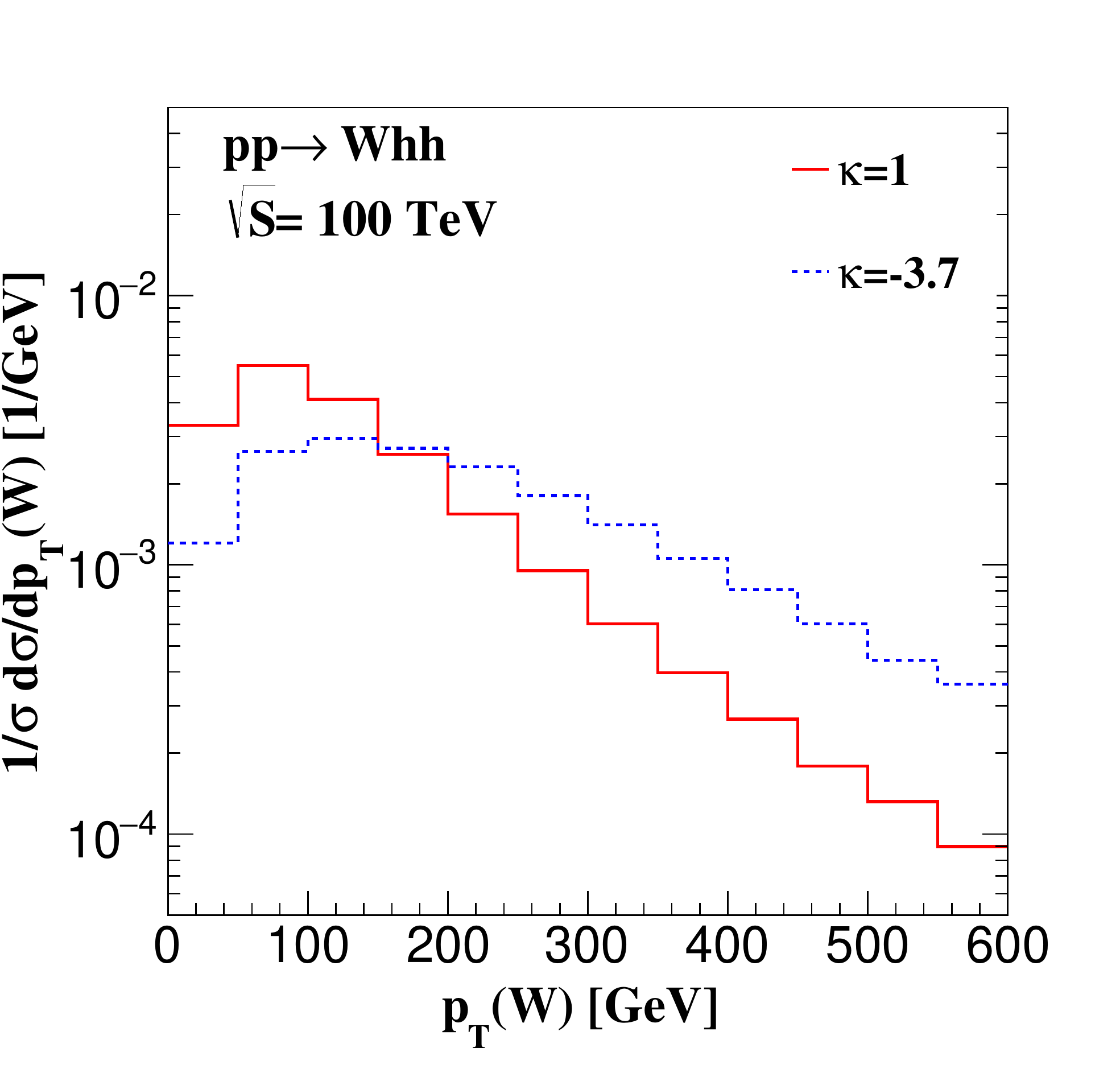}
    \includegraphics[width=0.45\linewidth]{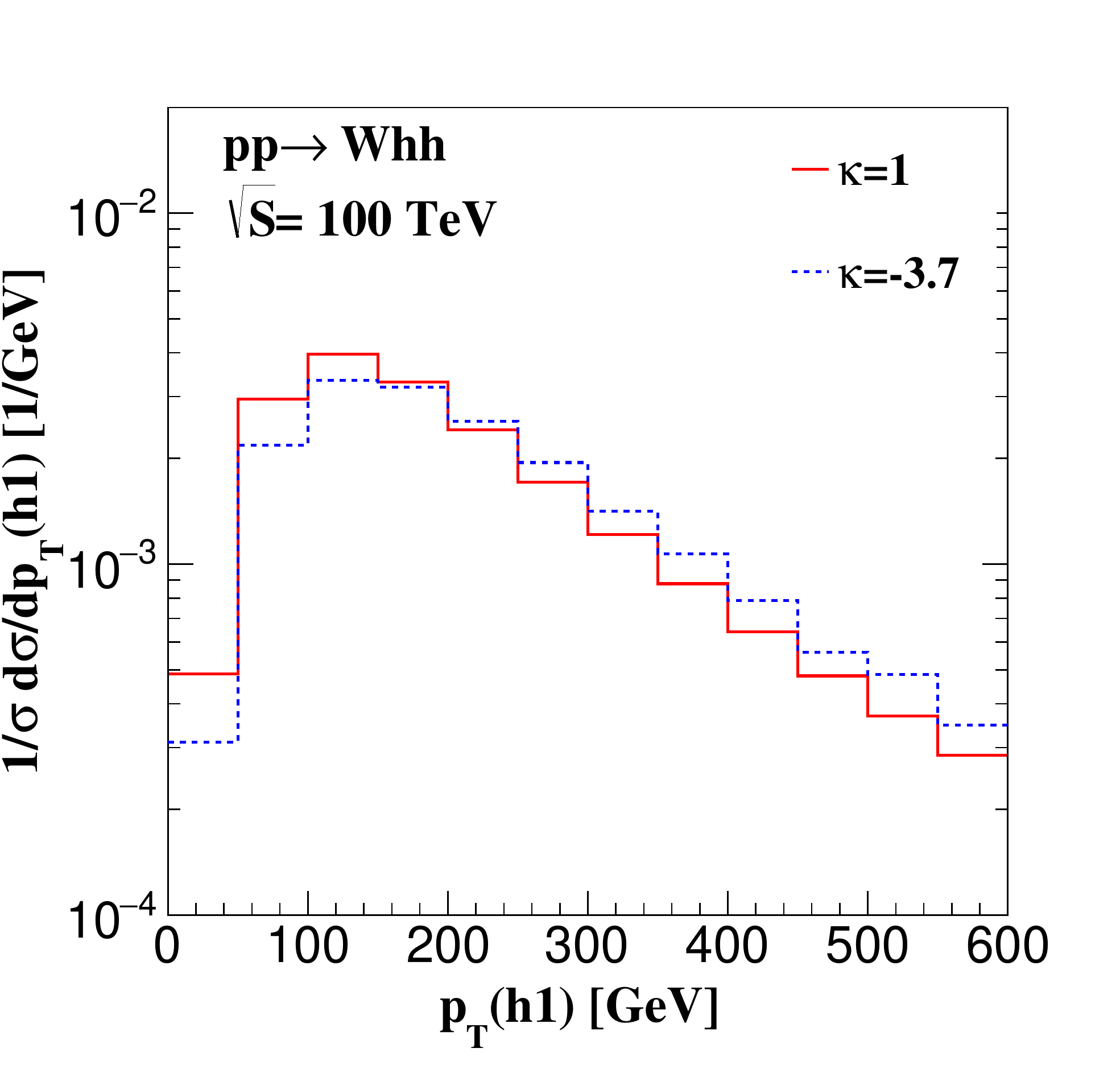}
  \caption{ Normalized NNLO $p_T$ distributions with $\kappa=1$ and $-3.7$, which correspond to the same total cross section. }
  \label{fig:kappa_dep_diff}
\end{figure}

Last but not the least, we investigate the sensitivity of this vector boson associated double Higgs production channel to the triple Higgs self-coupling.
In order to do so, we define a scaling factor $\kappa$ as
\begin{align}
\lambda_{hhh}=\kappa \lambda_{hhh}^{\rm SM}.
\end{align}
In Fig.~\ref{fig:kappa_dep}, we report the total cross section as a function of $\kappa$ at a 100 TeV hadron collider.
It can be seen that the total cross section is sensitive to the Higgs self-coupling.
Both the NLO and NNLO corrections are almost a constant for different values of $\kappa$.
However, for a given cross section, there are generally two values of $\kappa$, except for the minimum.
In Fig.~\ref{fig:kappa_dep_diff}, we present the normalized NNLO $p_T$ distributions of the $W$ boson and the leading Higgs boson with different self-couplings, $\kappa=1$ and $-3.7$, corresponding to the same total cross section.
At large $p_T$ region, the distributions of the $W$ boson and the Higgs boson are larger for $\kappa=-3.7$.
It means that one needs to investigate not only the total cross sections but also the kinematic distributions in order to determine the self-couplings.

\emph{Conclusions}:
It is crucial to measure the Higgs self-couplings after its discovery
in order to clarify the origin of electroweak symmetry breaking.
This can be achieved by studying the Higgs pair production at colliders.
The vector boson associated production plays a special role because
the lepton from the vector boson provides a clear tag of the event.
We present the QCD NNLO predictions on the total cross section
as well as the various kinematic distributions of this process.
The NNLO effects reduce the scale uncertainties,
and are sizable in the large transverse momentum region.
Then, we investigate the higher-order QCD effects in the presence of a jet veto.
They turn out to decrease the cross sections significantly.
The NLO cross sections suffer from large scale uncertainties
but the NNLO results are more stable.
We also study the sensitivity of this process to the Higgs self-couplings,
and find that the total cross section alone is not able to pin down the Higgs self-couplings.
The kinematic distributions are needed in order to achieve this goal.
These theoretical results can be utilized in future experimental analysis.

\emph{Acknowledgements}:
We would like to thank  Thomas L\"ubbert for helpful communication.
Parts of the computations were conducted on the
Mogon  cluster at Johannes Gutenberg University Mainz.
HTL is supported by the ARC Centre of Excellence for Particle Physics at the Tera-scale.
JW is supported in part by the Cluster of Excellence
{\it Precision Physics, Fundamental Interactions and Structure of Matter} (Grant No. PRISMA-EXC 1098).

\bibliographystyle{h-physrev}
\bibliography{whh}

\end{document}